\documentstyle[12pt]{article}
\begin{document}
\setcounter{secnumdepth}{1}
\title{A simple model for the formation of 
vegetated dunes}
\author{Hiraku Nishimori \\
        Hirohisa Tanaka\\
{\small \sl Department of Mathematical Sciences}\\
{\small \sl Osaka Prefecture University,Sakai 599-8531,Japan}\\
}
\maketitle

\bigskip
\noindent
{\large\bf ABSTRACT}
\bigskip
\noindent

A simple model for the dynamics of dunes associated with vegetation
is proposed. Using the model, formation processes of
transverse dunes, parabolic dunes and elongated parabolic dunes 
according to two environmental factors: 
i)the amount of sand at the source,ii)the wind force, 
are simulated.  
The results have qualitative correspondence to the real
counterpart, and the simplicity of the algorithm and the consequent 
easiness of the handling of this model provide us with wide applicability 
for the investigation of the complex interplay between 
vegetation and dunes.

\bigskip
\noindent
{\large\bf INTRODUCTION}
\bigskip
\noindent

It is  well known that 
the shape of dunes varies depending on 
several environmental factors surrounding 
them, like, the amount of available sand in each desert area, 
the wind directional variability, the vegetational condition covering 
sand surface, etc.
Among all, the interplay between the dynamics of dunes and the growth of 
plants remains many aspects unsolved because of the complex nature of 
the dynamics of the system and the difficulty of the inspection 
of the theoretical hyposesis through direct observations.
Still, many valuable 
observational studies have been accumulated which 
provided us detailed environmental conditions at 
individual arid areas in which 
typical types of vegetated dunes are seen\cite{Co,Hack,Th,Py,Ts,Ha}. 
One of the pioneering studies among them was made by Hack\cite{Hack}
in which he introduced a phase diagram to show the 
relation between the wind condition, available sand, the ratio of surface covered by plants, and the 
dominant type of dunes observed in each arid (or semiarid) area.
Here, taking previous observational studies into consideration,
we propose a minimal model which realizes the qualitative
dynamics of dunes in mildly vegetated arid areas.
Thereafter, using the model, the formation processes  of 
transverse dunes, parabolic dunes and elongate parabolic dunes
according to environmental factors are simulated, that is,
the Hack's phase diagram of vegetated dunes is numerically 
testified.
In the below we firstly
make a brief explanation of our model 
and, after that, the outputs obtained through the simulation are shown.
Finally we make a discussion on the meaning of the model.

\bigskip
\noindent
{\large\bf MODEL}
\bigskip
\noindent

The model is a 2-dimensional(2D) lattice model, wherein 
two field variables,i)the local height $h(i,j,n)$ of sand bed, and
ii)the local density of vegetation $c(i,j,n)$, are allocated at each site of 
the horizontally extending $N\times N  (N=100)$ lattice.
Here $\{(i,j)|1\le i,j\le N\}$ and $\{n |0\le n\}$ indicate,respectively, horizontal position and time step. 
Assume wind is constantly blowing in positive $i$ direction, and 
each $(i,j)$ site covers a sufficiently wider area than that 
occupied by individual grains. Likewise is the unit time step  
which reflects a much longer time than individual saltation 
processes of sand.  
The above field variables are set to 
interact each other through the suppression factor $a_\beta(i,j,n)$
as explained below.  
Although many other quantities are considered to contribute to the
whole dynamics of the system, 
we concentrate ourselves 
on extracting a simple set of relevant factors to investigate the essential 
dynamics of the whole system. 
Such methodology for modeling the complex dynamics of dunes with a 
set of simple rules have recently been introduced 
by Warner\cite{We} and also by one of present authors\cite{Ni}, 
with which they 
succeeded to numerically reproduce the typical types of unvegetated dunes under conditions qualitatively corresponding to 
the observations made by Wasson and Hyde\cite{Was}. 
More previously, a pioneering model of the formation of vegetated dunes
using a minimal set of rules has been introduced by Castro\cite{Ca}.
Although his model is not applicable for the dynamics of dunes without vegetation, neither, does reproduce a wide variety of morphology of dunes in 
vegetated areas, it successfully simulated a systematic change of the inter-dunes length 
of vegetated transverse dunes, also the change of their profiles.
 
Present model is a hybrid model to investigate a wide class 
of morphodynamics of dunes with vegetation also without vegetation.
Here, dynamical rules of sand movement are based on a 
previous work of the author\cite{Ni}, 
whereas the rules for time evolution of 
plants have many similarities to the model by Castro\cite{Ca}.

\noindent
I. For the evolution of $c(i,j,n)$, the local density of plants,
we adopt a set of simple rules.
In an extreme case where the height of sand surface remains unchanged 
with time, 
wherein plants can grow thick up to the saturation density without 
being cut away by the drastic deflation of ground or being    
buried by the rapid accumulation of sands,
$c(i,j,n)$ is allowed to increase linearly until the maximum value $c_{max}$.
On the other hand, if the temporal change of surface height is too
fast, the growth rate of plants is suppressed or some of them may 
wither up and die, then $c(i,j,n)$ decrease down to the minimum limit $c_{min}$.
To reflect the above situation, we use a discrete dynamics 
which is a sectional linear map as shown in fig.1. 
Specifically, the dynamics is expressed as, 
\begin{eqnarray*}
c(i,j,n+1)&=& A(c(i,j,n)-b(i,j,n))+c_{mim} \qquad (b \le c \le (c_{max}-c_{min})/A +b) \\
c(i,j,n+1) &=& c_{max} \qquad ((c_{max}-c_{min})/A +b < c )\\
c(i,j,n+1) &=& c_{mim} \qquad (c < b )\\
\end{eqnarray*}
here $b \equiv b(i,j,n) \equiv |h(i,j,n)-h(i,j,n-1)|$,
$c \equiv c(i,j,n)$ and $A$ is a constant to determine 
the growth rate of plants.

\noindent
II. for the evolution of $h(i,j,n)$,
discretized conservation law of $h(i,j,n)$,
\begin{equation}
h(i,j,n+1)-h(i,j,n) =  Q_{in}(i,j,n)-Q_{out}(i,j,n)
\end{equation}
holds, where $Q_{in}(i,j,n)$ is the total mass of sand
coming into site $(i,j)$ at time step $n$, while $Q_{out}(i,j,n)$ is the 
same quality leaving from $(i,j)$ at $n$. 
Both of the saltation flux and the creep flux contribute 
to $Q_{in}(i,j,n)$ and $Q_{out}(i,j,n)$.
Specifically the saltation flux caused by the grains 
leaving from $(i,j)$ at $n$ is expressed  
by the production of its mass $m$ and length $l$ like, 
\begin{equation}
q_{sal}(i,j,n)=q_0(tanh(\nabla h(i,j,n))+1.0)(tanh(-\nabla h(i,j,n))+1.0+\alpha).
\end{equation}
where $\nabla h(i,j,n)$ means $h(i,j,n)-h(i-1,j,n)$ and
$\alpha$ is a constant to determine the bed-load in the 
windward slope of dunes.
The above reflects the qualitative nature of wind, 
also the resulting saltation flux around dunes which is 
intensive in the windward particularly around  
the crest, whereas almost no flux in the lee side\cite{Ni,Ras}(fig.1(b)). 
On the other hand, the flux by creep $q_{creep}(i,j,n)$ is set 
proportional to the local 
gradient of sand surface. Although this may be a very crude approximation,
it has some sense for a qualitative description of the morphodynamics of dunes
as discussed afterward.
The crucial effect caused by permitting the growth of plants is 
such that the sand transport sharply decreases when
the cover ratio of sand surface by plants exceeds  
a critical value.
To realize the situation in  a simple expression,
the suppression factor $a_\beta(i,j,n)$ is introduced like,
\begin{equation}
a_\beta(i,j,n)=1+\beta(tanh(c-c_{cr})-1). \qquad (0 \leq \beta \leq 1/2 ) 
\end{equation}
Here, $c_{cr}$ is the critical vegetation density 
over which the movement of sand sharply decreases, and $\beta$ 
determines the maximum efficiency of suppression(fig.1(c)).
The value of $\beta$ depends on whether it is for 
the saltation flux or for the creep flux.  
With the suppression factor, 
the saltation/creep flux is forced to decrease as 
$q(i,j,n)=a_{\beta}(i,j,n)^2 q'(i,j,n)$
where $q(i,j,n)$ and $q'(i,j,n)$ are, respectively, the local flux by saltation/creep with vegetation and without vegetation,.

\bigskip
\noindent
{\large\bf RESULTS}
\bigskip
\noindent

Intending to compare the simulation outputs with the diagram by Hack, 
two kind of quantities are chosen as the control parameters;
One is i)the amount of sand at the source.
Namely, the average height, $<h_{source}>$, of the sand surface
at the source sites, $\{(i,j)|i=1,2 \quad 1\le j\le N \}$. 
Specifically, uniformly random numbers between 
2$<h_{source}>$ and 0 are allocated on these sites at each 
time step. The other control parameter is
ii)the wind strength which should be a monotonically increasing function 
of saltation flux. Specifically, the variable $q_0$ in eq.(2) 
is adopted as the index of wind strength.  
Note that the vegetation density, which is one of the axes in  
the diagram by Hack,
is not adopted as a control parameter 
because it is rather a resultantly attained quantity after 
the above two control parameters are fixed.
Simulations are initiated from flat sand surface except the source area, 
namely, $h(i,j,0)=0$  holds except the source sites, 
while the initial vegetation density
$c(i,j,0)$ at each site is set random around the 
average value $<c(i,j,0)>$ which is between $c_{max}$ and $c_{min}$.
Note the boundary condition in $j$ direction, which is perpendicular to 
the wind direction, is set periodic and the leeward boundary in $i$ 
direction 
is set as the free boundary.
Below the initial level 0 of sand surface, 
erosion of sand surface is inhibited to realize the existence 
of the hard ground 
or the ground water table.
Using these rules, 
spontaneous formation processes of typical types of dunes 
are observed.

The results are; 
First of all, when the wind force is too weak, regardless of
the amount of sand supply at the source, clearly shaped dune will
not appear in the system(fig.2(a)).
It is also the cases where the amount of sand supply is too small.
On the other hand, with more than a certain strength of wind force
and a certain amount of sand supply, 
two types of clearly shaped active dunes are formed;

\noindent
i)When sufficiently large amount of sand is supplied under
rather strong wind, transverse dunes, barchan dunes 
 or both of them will dominate in the system. 
In more detail, small parabolic dunes  formed 
just lee of the sand source soon develop into barchan dunes, 
which connect each other to grow larger as they move, to form transverse dunes
the crests of them extend, roughly, perpendicular to the direction of 
wind(fig.2(b)). 

\noindent
ii)If the amount of supplied sand or the wind force is slightly less than 
the above, parabolic dunes will prevail in the system(fig.2(c)(d)).
They have arms extending to the windward direction.
At the center of each parabolic dune, hollow is developed 
in which surface erosion proceeds down to the level of 
hard ground to serves the dune's nose with sand.
There we can see the inclination that the length of the arms varies depending 
on the wind force, namely, the arms extend the longer   
under the stronger wind. Especially in the case 
with rather small amount of sand, elongate parabolic dunes\cite{Py} with  
the shapes like  hair-pins will appear(fig.2(d)).  
The arms of such elongate dunes, if without their noses, 
seem like pairs of linear dunes. 

\bigskip
\noindent
{\large\bf DISCUSSION}
\bigskip
\noindent

The above results are summarized in a phase diagram as shown in fig.4.
This diagram is not directly compared with that of Hack 
because of the less number of axes in our diagram.
Also, in the diagram, ambiguous area is left 
in which condition irregular mounds of sand are formed, 
which are not easily categorized into  
other dominant types of dunes.
Still, it has qualitatively good correspondence 
to the previous observational studies including that of Hack. 
Especially, the systematic change of the dunes' morphology from 
the transverse(or barchan) dunes to the parabolic or the elongate 
parabolic dunes according to 
the amount of available sand (or to the wind force), is clearly simulated. 
Moreover no dune formation is realized under too weak wind nor 
is under too small amount of sand supply.
The results indicate that this simple model contains 
intrinsic dynamics by which  
the morphodynamics of the vegetated dunes 
is decisively affected, and that through this model theoretical 
hypotheses on the formation dynamics of
vegetated dunes can be testified. 

On the other hand, present study leaves   
several, possibly crucial, situations untreated.
Firstly, the wind is always blowing in one direction.
This condition seems very hard for the spontaneous growth of linear dunes
considering the cases of unvegetated dunes. 
Consequently, straightly extending sand dunes/ridges  
appear only in behind parabolic dunes as the traces of their 
arms. This is one of the possible 
origin of vegetated linear dunes\cite{Py}. 
But it seems not the general scenario for the formation of
linear dunes, thus, the study under complicated wind regimes
is one the next issues.
Secondly, the existence of the angle of repose 
was not explicitly incorporated in the model.
Here, instead, all the dynamics along sand surface 
was modeled by the diffusion-like creep process.
Of course such a rough simplification can cause 
the imperfect reproduction of the real counterparts, like  
inhomogeneous angle of slope at the lee face of a dune unlike the 
actual slip face. 
On the problem, in the previous studied, we demonstrated that
even such a simplified modeling can work effectively, 
at least for the investigation of  
macroscopic morphodynamics of unvegetated dunes, 
like, simulating distinguishable barchan, seif, star and other types of dunes.
Also in the present simulations with vegetation,
distinguishable parabolic and other types of dunes  
successfully appeared according to the corresponding 
situations in the semiarid desert areas.
However, for the investigation of more quantitative aspects of the 
system, the explicit introduction of the angle of repose into the model is
required, and it also be an important step to the inclusive understanding 
of the system.

Notwithstanding all, we believe the present study using such a 
minimal model is one of the effective way to understand the complex 
interplay between dunes and vegetation. Systematic search
of complex systems using such bold simplifications would , more or less 
work as the complementary way which fill the big blank between 
the observational studies of such systems 
and the, otherwise hardly testified, theoretical hyposesis. 

\bigskip
\noindent
{\large\bf ACKNOWLEDGEMENTS}

\bigskip
\noindent
This research was supported by the Grant-in-Aid for Scientific Research
of JSPS (c11837017).
 
\pagebreak

\pagebreak

\bigskip
\noindent
{\Large\bf Figure Captions}

\vspace{.2in}

\noindent
Fig.1  Schematic explanations of the present model;
(a-I)The Map $c(i,j,n) \rightarrow c(i,j,n+1)$ 
to describe the discretized time evolution of 
plants density. Without the temporal change $b(i,j,n)$
of surface height, plants density monotonically 
increases up to the saturation value $c_{max}$.
(a-II)With more than a certain speed of surface rise or deflation,
plants at the surface are, more or less, damaged because they are buried or 
cut away, then, $c(i,j,n)$ will decrease with time 
with the lower limit $c_{min}$. 
b)Local sand flux by saltation $q_{sal}$ of eq.(2) is given as a function of local slope $\nabla h$ of sand surface in the direction of wind, 
which reflects the observational fact,
namely, large bed load in the windward particularly around  
the crest(indicated by the arrow), whereas sharply it decreases in the lee side\cite{Ni,Ras}.
c)Suppression factor $a$ of sand flux as a function of local 
density of vegetation $c$. Above a critical density $c_{cr}$ of vegetation, 
the flux of sand drastically decreases as described in eq.(3).

\bigskip
\noindent
Fig.2  Snapshots of simulated dunes under various pairs of
control parameters, wind force and sand supply.
The left part in each figure shows the spatial distribution of vegetation
density $d(i,j,n)$.The darker tone indicates the more densely vegetated place,
whereas white parts indicates the areas with bare sand surfaces. 
The right part in each figure shows the surface height distribution 
$h(i,j.n)$, where the darker position means the higher surface.
Sand is supplied from the most windward 2 rows. 
Steady wind is blowing from the left to the right.
(a)When the wind force is too weak no distinguishable dune
is formed. It is also the case for too small amount of sand supply
at the source. 
(b)Under rather strong wind with sufficient amount of sand supply
transverse dunes prevail while 
small parabolic dunes are seen just lee of the sand source.
The latter will soon grow up to the former.
(c)Parabolic dunes, the arms of which 
extend in the windward direction, are formed 
under mildly blowing wind with intermediate amount of sand at the source.
(d)If the amount of sand is comparatively small within this regime,
thin and long parabolic dunes, namely, elongate parabolic dunes will grow. 
They look like rather linear dunes if without their noses. 

\bigskip
\noindent 
Fig.3 Phase diagram to show the dominant types of dunes 
under various pairs of control parameters. 
The alphabets in the diagram indicate the conditions 
corresponding to respective snapshots in fig.2.
Symbols $\triangle$s and $\bullet$s indicate, respectively,
parabolic dunes, and, barchan(or transverse) dunes, whereas 
$\times$s mean the conditions for no dune formation.
The $\triangle$s accompanied by {*} mean the conditions for
the development of rather thin parabolic dunes with long arms. 
At the condition with the symble $\Box$, many irregular mounds are 
formed which are not clearly categorized as particular type of dunes.


\addcontentsline{toc}{section}{REFERENCES CITED }

\begin{thebibliography}{999}

\bibitem{Co}
Cook,R.,Warren,A., and Goudie, A., 1993, Desert Geomorphology: London, UCL press.

\bibitem{Hack}
Hack,J.T., 1941, Dunes of the western Navajo Country: Geographical Review,
 v.31, p.240-263.

\bibitem{Th}
Thomas,D.S.G., and  Tsoar,H., 1990, The geomorphological role of vegetation in 
 desert dune systems: Vegetation and Erosion, p.471-489.

\bibitem{Py}
Pye,K., 1982, Morphological development of coastal dunes in a humid
 tropical environment, cape bedford and cape flattery, North Queensland: 
 Geografiska Annaler, 64A, p.213-227.

\bibitem{Ts}
Tsoar,H., and Moller,J-T., 1982, Aeorian Geomorphology:Boston, Allen and Unwin, p.75-97.

\bibitem{Ha}
Halsey,L.A., Catto,N.R., and Rutter,N.W., 1990,
Sedimentology and development of parabolic dunes, Grande Prairie dune
 field, Alberta: Canadian Journal of Earth Sciences, v.27, p.1762-1772.

\bibitem{We}
Werner,B.T., 1995, Eolian dunes: Computer simulations and attractor
 interpretation:Geology. v.23, p.1107-1110.

\bibitem{Ni}
Nishimori, H., Yamasaki, M., and Anderson, K. H., 1998, A simple model for the
 various pattern dynamics of dunes: International Journal of Modern
 Physics, v.12, p.257-272.

\bibitem{Was}
Wasson,R.J.,and Hyde,R., 1983, Factors determining desert dune type
:Nature,v.304, p.337-339.

\bibitem{Ca}
de Castro,F., 1995, Computer simulation of the dynamics of a dune
 system: Ecological Modelling, v.78, p.205-217.


\bibitem{Ras}
Rassmussen,K., 1989, Proc of Symposium 'coastal sand dunes':Edinburg,
 Royal Society of Edinburg, B96, p.129-147.


\end{thebibliography}
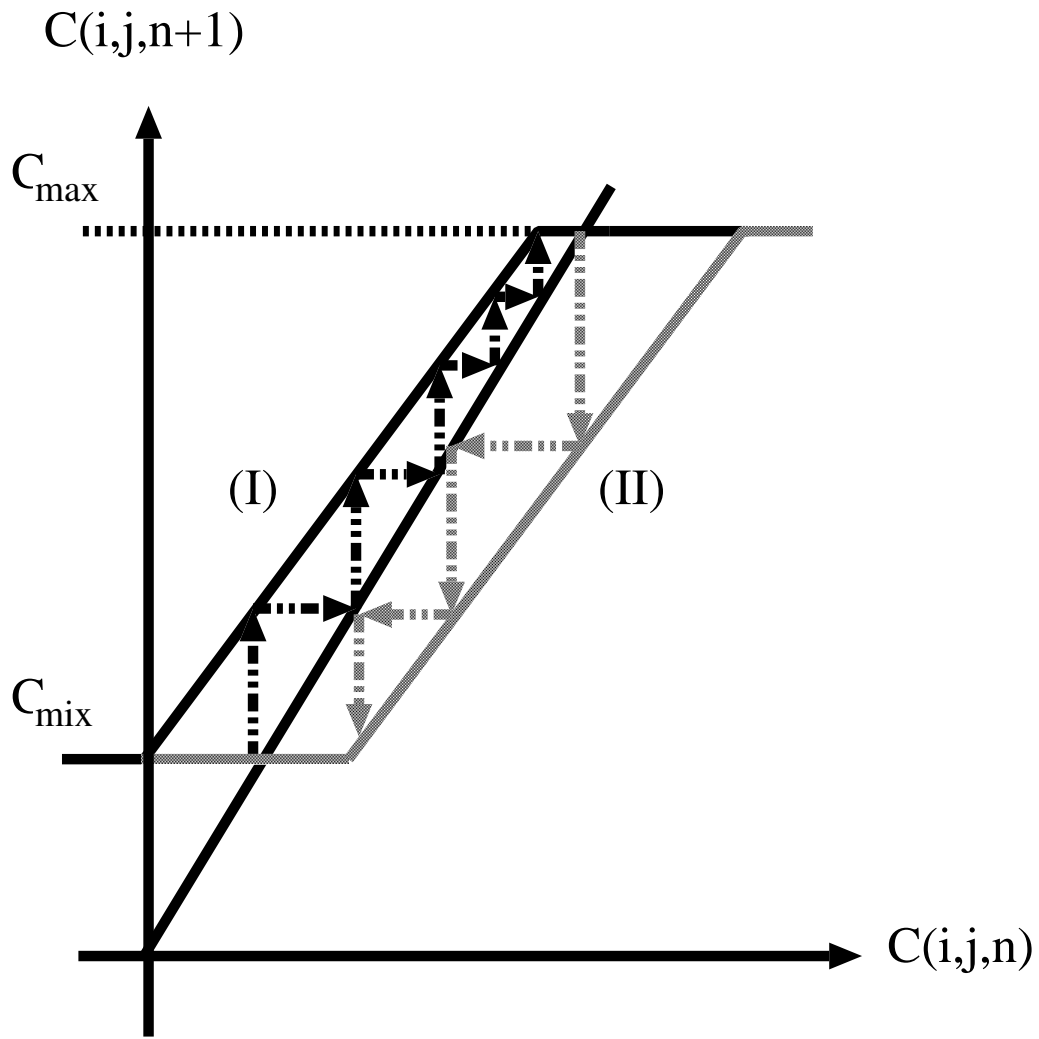
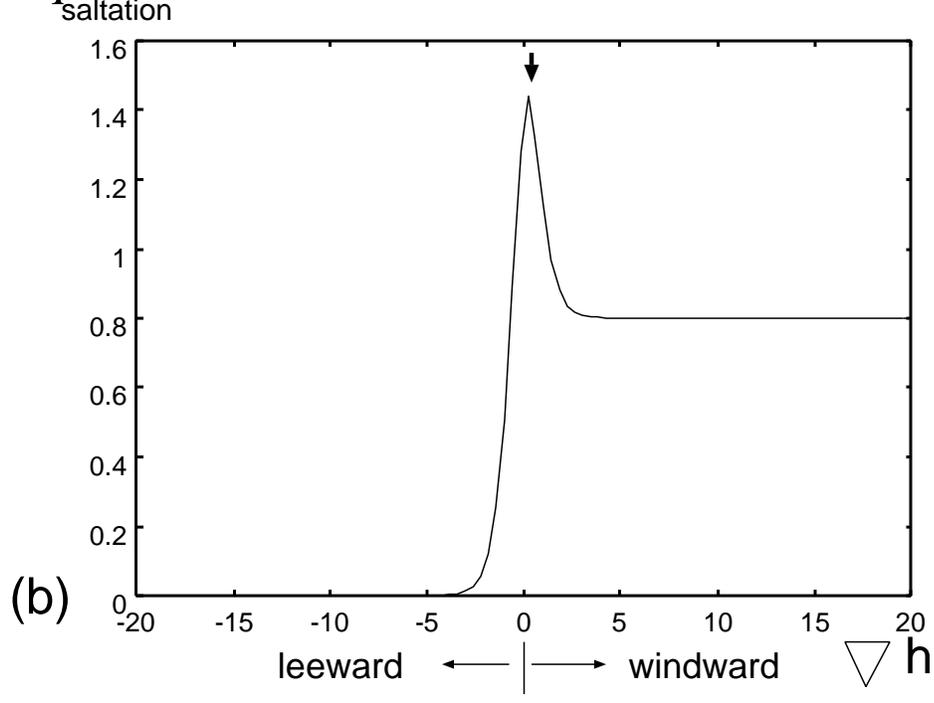
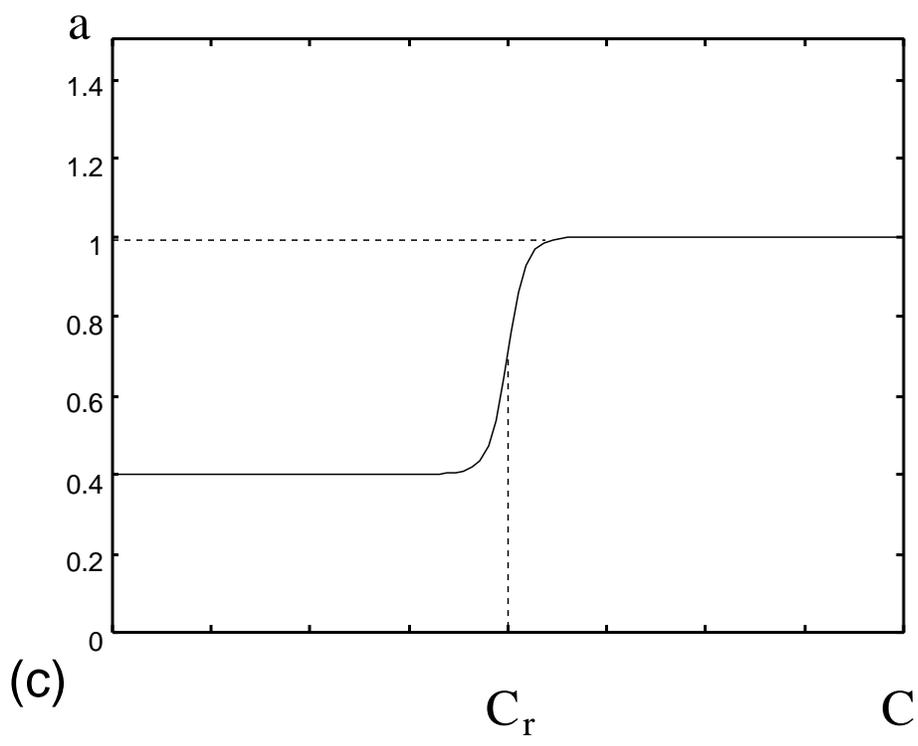
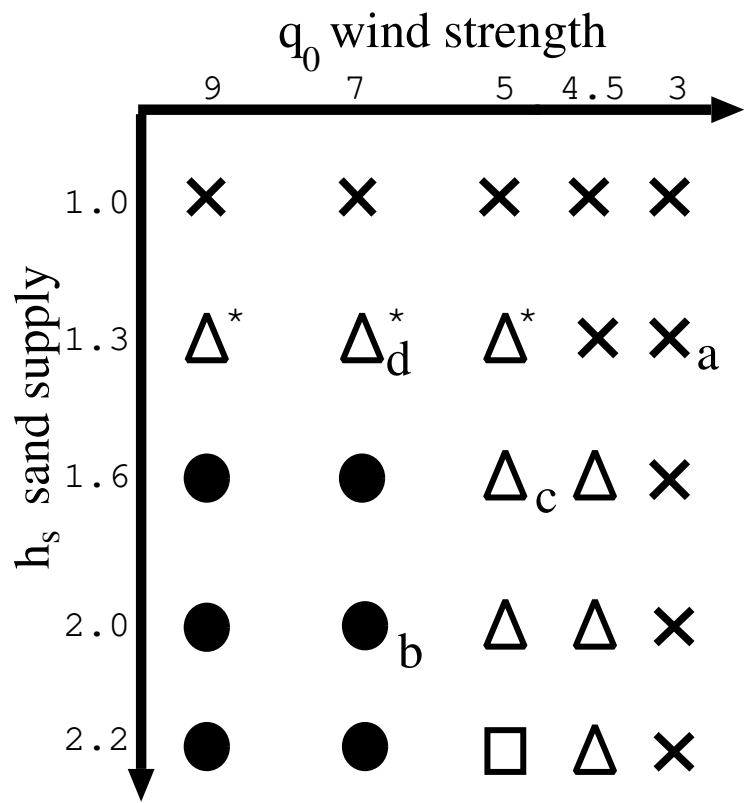
\end{document}